\documentclass[sigconf,nonacm]{acmart}

\usepackage{balance}

\AtBeginDocument{%
  }

\begin{document}

\title{GRaSP: Latency-Aware Optimization for\\ Geo-Distributed Retrieval-Augmented Generation}
\thanks{Accepted at SIGIR-AP 2026, December 13--15, 2026, Singapore.
The definitive version is available at \url{https://doi.org/10.1145/3848000.3850202}.}

\author{Hardik Ruparel}
\orcid{0009-0000-4805-1227}
\affiliation{%
  \institution{Independent Researcher}
  \city{San Jose}
  \state{CA}
  \country{USA}}
\email{hruparel@uci.edu}

\author{Raj Kumar Pandey}
\orcid{0009-0008-4553-3018}
\affiliation{%
  \institution{The Pennsylvania State University}
  \city{University Park}
  \state{PA}
  \country{USA}}
\email{rzp5412@psu.edu}

\author{Pramod Kumar}
\orcid{0009-0000-7426-367X}
\affiliation{%
  \institution{Independent Researcher}
  \city{Sunnyvale}
  \state{CA}
  \country{USA}}
\email{pramodkumar5878@gmail.com}

\renewcommand{\shortauthors}{Ruparel, Pandey, and Kumar}

\begin{abstract}
Retrieval-Augmented Generation (RAG) is increasingly served across geo-distributed cloud and edge infrastructures, where wide-area latency, cross-region egress cost, and data-residency rules compete with retrieval quality. Conventional pipelines rank by semantic similarity alone and ignore \emph{where} content lives, inflating tail latency and cost. We formalize the \emph{Geo-RAG Selection Problem} (GRaSP), a drop-in re-scoring layer that decays relevance by latency and adds an optional cost penalty under a cardinality budget and per-region caps, with no retraining. We first settle its complexity: the base formulation is a linear objective over a partition matroid, solved \emph{optimally} by an $O(N\log N)$ greedy algorithm, rather than NP-hard as a knapsack framing suggests, with hardness entering only under an additive egress budget. On a five-node CloudLab testbed emulating a four-region WAN over the Natural Questions corpus (2.68M passages), GRaSP's benefit is \emph{placement-dependent}: under uniform placement it cuts fetch latency at negligible quality cost; under topical (residency-like) placement it traces a recall--latency Pareto frontier that dominates relevance-only, nearest-$k$, and a latency-budgeted baseline. At a moderate latency setting, GRaSP lowers P90 fetch latency by $41\%$ while retaining $78\%$ of relevance-only's recall, with sub-millisecond selection overhead, stable to 200~queries/s. Findings generalize across four public BEIR benchmarks. In end-to-end question answering, GRaSP reduces context-fetch latency by $43\%$ while remaining within $0.062$ absolute EM of relevance-only retrieval.
\end{abstract}


\maketitle

\section{Introduction}
Large language models (LLMs) excel at generating text, but they can hallucinate, and their knowledge can become outdated. Retrieval-Augmented Generation (RAG) helps by retrieving relevant passages at inference time and giving them to the model as additional context~\cite{lewis2020rag}. RAG is now widely used in enterprise assistants, developer copilots, and analytics systems. 
In practice, the data used by these applications is often spread across multiple cloud regions and edge locations. Organizations do this to reduce client latency, control costs, improve availability, and meet data-residency requirements. This creates a new challenge for RAG: the most relevant passage is not always the best one to fetch. A highly relevant passage may be stored in a distant region, making it slower and more expensive to retrieve. Regional policies may also limit how much content can be fetched from a particular location. As a result, relevance-only top-$k$ selection can increase tail latency, raise egress cost, and behave poorly under regional constraints.

Consider a SaaS assistant whose knowledge base is distributed across \texttt{eu-central}, \texttt{ap-southeast}, and \texttt{sa-east}. A user in Paris is served from \texttt{eu-central}, but several of the most relevant passages for a query are stored in \texttt{sa-east}, roughly 115\,ms away. A relevance-only policy will select passages without considering their storage locations, increasing both fetch latency and cross-region egress costs.
If similar passages are available in Europe, choosing them instead can keep most of the retrieval quality while reducing latency. But if the useful content exists mainly in \texttt{sa-east}, then avoiding the remote fetch will also reduce relevance. In other words, the benefit of location-aware retrieval depends on \emph{where relevant content is placed}.

We formalize this problem as the \emph{Geo-RAG Selection Problem} (GRaSP). GRaSP runs after retrieval or re-ranking and before the selected passages are fetched. It re-scores candidates using semantic relevance, network latency, and optional retrieval cost, while also enforcing global and per-region selection limits. Because GRaSP operates only on the candidate pool produced by the existing retriever, it can be added to dense or hybrid RAG pipelines without retraining or re-indexing.
Although the objective includes latency decay, cost, and regional constraints, the base optimization problem is simple. For a given query, each candidate has a fixed utility. This reduces the problem to a linear selection problem that can be solved exactly using an $O(N\log N)$ greedy algorithm. The problem becomes NP-hard only when we add constraints such as a hard egress budget.

Our evaluation shows that GRaSP's benefit depends strongly on data placement. When relevant content is spread across regions or well replicated, GRaSP can reduce latency and cost with little loss in retrieval quality. When relevant content is concentrated in a remote region, latency and relevance are in direct tension. In that case, GRaSP provides a tunable trade-off rather than eliminating the cost of remote access.

Replication can reduce this tension. By placing copies of frequently accessed content in a small number of additional regions, the system creates more nearby alternatives for GRaSP to choose from. Our results show that this can recover much of the low-latency behavior seen under uniform placement. The main contribution of this work is therefore not a complex optimization algorithm, but a clear formulation and an empirical study of when geo-aware retrieval helps. Specifically:

\begin{itemize}
\item We formulate GRaSP as latency-aware relevance maximization with an optional cost penalty, a global top-$k$ limit, and per-region caps (Section~\ref{sec:formulation}). GRaSP is a drop-in re-scoring layer and requires no retraining or re-indexing.

\item We show that the base problem can be solved optimally by an O(NlogN) greedy algorithm, while NP-hardness arises only when additive resource constraints, such as a hard egress budget, are introduced (Section~\ref{sec:complexity}).

\item We build a five-node CloudLab~\cite{CloudLab} testbed that emulates a four-region WAN and evaluate GRaSP on 2.68M Natural Questions passages. We measure retrieval quality, tail latency, egress cost, and optimality against an exact solver (Section~\ref{sec:eval}).

\item We show that GRaSP's benefit depends on data placement, and that replication can improve the quality--latency trade-off. We further show that latency ($\gamma_q$) and cost ($\beta$) can be tuned independently, that these findings generalize across four BEIR datasets, and that GRaSP incurs only sub-millisecond overhead while remaining stable at query rates of up to 200~queries/s.

\end{itemize}
\section{Related Work}
\textbf{RAG and Adaptive Retrieval:} RAG research has emphasized retrieval quality and fusion strategies~\cite{lewis2020rag}, along with graph-structured retrieval~\cite{peng2024graphrag} and, more recently, distributed RAG architectures~\cite{xu2025drag}. Adaptive and cascaded retrieval methods decide \emph{how much} to retrieve or \emph{when} to stop~\cite{jiang2023active}, but they optimize over a centralized index and do not treat network location as a selection signal. GRaSP is orthogonal and composable: it re-scores whatever candidate pool an upstream retriever or re-ranker produces.

\textbf{Geo-distributed Analytics and Storage:} A large systems literature optimizes data and task placement across wide-area networks for latency and cost. Iridium~\cite{pu2015iridium} jointly optimizes data and task placement to reduce wide-area query latency; Uluyol et al.~\cite{uluyol2020tradeoff} quantify latency versus cost trade-offs in geo-distributed storage, and transaction chains~\cite{zhang2013chains} achieve low-latency serializability. SPANStore~\cite{spanstore} minimizes the cost of geo-replicated storage placement across multiple clouds. These works decide \emph{where to place or how to replicate} fixed objects; GRaSP instead decides \emph{which} objects to fetch at query time given a fixed placement and uncertain relevance, and is complementary to placement optimizers.

\textbf{Caching and Content Placement:} Content caching in CDNs and edge networks decides \emph{which} content to keep close to users to reduce latency; recent work makes this decision \emph{cost-aware}, accounting for the external bandwidth and retrieval cost of cache misses rather than hit-ratio alone~\cite{araldo2014costcache,wang2021cost}. In the LLM setting, semantic caches reuse prior responses or retrieved contexts for similar queries to cut inference cost and latency~\cite{gptcache, ruparel2025ragcachesim}, and in multi-tenant RAG, frequency-aware global caches outperform per-tenant partitioning under skewed workloads~\cite{ruparel2025caching}. These mechanisms operate \emph{ahead of} a query, provisioning content by predicted popularity; GRaSP instead operates \emph{at} query time, selecting among already-placed candidates under per-query relevance, latency, and cost. The two are complementary---our replication result (Section~\ref{sec:replication}) shows that caching hot content into nearby regions widens the regime in which GRaSP's selection is near-free.

\textbf{Replica and Server Selection:} Systems such as DONAR~\cite{wendell2010donar} and C3~\cite{suresh2015c3} choose \emph{which replica} of a known object to contact to minimize latency under load. The key distinction is that replica selection assumes the target object is fixed and replicated, whereas GRaSP must jointly decide \emph{which} documents to include---trading semantic relevance against location---before any replica choice is made. When documents are replicated, GRaSP naturally subsumes best-replica selection as the inner fetch decision (Section~\ref{sec:replication}).

\textbf{Constrained Selection:} Algorithmically, GRaSP's cap-constrained selection is a linear objective over a partition matroid; its budgeted variant relates to the 0--1 knapsack~\cite{karp2009reducibility,martello1990knapsack}. We make this structure explicit in Section~\ref{sec:complexity}, which clarifies when exact selection is cheap and when approximation is warranted. GRaSP is also complementary to federated and decentralized knowledge access~\cite{koneccny2016federated}.

\section{Problem Formulation}
\label{sec:formulation}
Consider a query $q$ originating from a source region $s$. Let
$\mathcal{D} = \{d_1, \dots, d_N\}$ denote the set of candidate documents, where each document
$d_i$ is physically hosted in some region $r(i)$. We assume an upstream retriever or re-ranker provides a normalized semantic relevance score $\mathrm{rel}_i \in [0,1]$ for each $d_i$. We
associate two system-level attributes with every candidate:
\begin{itemize}
  \item \textbf{Network latency} $\mathrm{lat}_i(s)$: the estimated round-trip latency (ms)
  from source region $s$ to hosting region $r(i)$.
  \item \textbf{Retrieval cost} $\mathrm{cost}_i \ge 0$: the per-document egress or metered
  access charge for retrieving $d_i$ from its hosting region $r(i)$ to $s$.
\end{itemize}

We assume the retrieval index (document embeddings) is available at or near the source region, so scoring all candidates is local and inexpensive; the latency and egress costs we model are incurred only when the \emph{content} of a selected document is fetched from its hosting region for inclusion in the context window. This models a deployment in which compact index representations are replicated near serving regions while full passage payloads remain regionally partitioned. Consequently, $\mathrm{rel}_i$ is known for every document in the candidate
pool before any selected passage payload is fetched, and GRaSP operates strictly between scoring and fetching: it decides \emph{which} scored candidates justify their retrieval cost.

We introduce binary decision variables $x_i \in \{0,1\}$, where $x_i = 1$ means $d_i$ is
selected for the context window. The selection respects a global top-$k$ budget and per-region
capacity constraints:
\begin{align}
\sum_{i=1}^N x_i &\le k, \label{eq:cardinality}\\
\sum_{i:\, r(i) = r} x_i &\le C_r \quad \forall r, \label{eq:regional}
\end{align}
where $k$ is dictated by context-window/token limits and $C_r$ encodes regional balance or data-residency limits. Because \eqref{eq:cardinality} is an inequality, a selection may include fewer than $k$ documents when further retrievals would not improve the objective.

\subsection{Latency-decayed utility with cost penalty}
GRaSP selects documents to maximize a latency-aware objective while optionally penalizing
retrieval cost:
\begin{equation}
\label{eq:objective}
\max_{x \in \{0,1\}^N}
\ \sum_{i=1}^{N} x_i \left[ \mathrm{rel}_i \, e^{-\gamma_q\,\mathrm{lat}_i(s)} \right]
\;-\; \beta \sum_{i=1}^{N} x_i\,\mathrm{cost}_i,
\end{equation}
where $\gamma_q \ge 0$ is the latency-sensitivity parameter, set globally or per query (larger $\gamma_q$ penalizes
remote content more) and $\beta \ge 0$ is the cost-sensitivity weight. The term
$e^{-\gamma_q\,\mathrm{lat}_i(s)}$ decays a document's relevance in proportion to its latency
from $s$. Writing $u_i = \mathrm{rel}_i\,e^{-\gamma_q \mathrm{lat}_i(s)} - \beta\,\mathrm{cost}_i$ for each document's net utility, the objective is $\max_x \sum_i u_i x_i$, linear in the decision variables. We use an exponential decay because it is smooth, monotone, and scale-free in the product $\gamma_q\,\mathrm{lat}_i(s)$, and because it keeps each $u_i$ a precomputable constant.

\paragraph{Interpretability}
Each parameter has a direct operational meaning: $k$ limits total retrieved documents; $C_r$
enforces per-region balance or residency; $\mathrm{rel}_i$ is semantic relevance;
$\mathrm{lat}_i(s)$ is network delay; $\mathrm{cost}_i$ is monetary fetch cost; $\gamma_q$ tunes
the latency penalty (higher for interactive chat, lower for batch analytics); and $\beta$ tunes
the cost penalty. This lets practitioners adapt the model to low-latency chatbots,
cost-sensitive batch jobs, or compliance-critical deployments.

\paragraph{Additive latency as a surrogate}
Objective~\eqref{eq:objective} penalizes each selected document's latency
\emph{additively}, whereas parallel fetching makes a selection's realized fetch latency its
\emph{maximum} (Eq.~\ref{eq:lmax}), not its sum. The separable objective is thus a
serving-efficient surrogate for the non-separable maximum: penalizing every remote document
suppresses the worst-case fetch while keeping each coefficient $u_i$ precomputable and the problem
exactly greedy-solvable (Section~\ref{sec:complexity}). Section~\ref{sec:online} shows the
surrogate tracks measured end-to-end latency closely; directly optimizing $\bar L_{\max}$ is also
polynomial via a latency-threshold sweep, a natural extension.
\section{Complexity of Geo-RAG Selection}
\label{sec:complexity}

GRaSP may initially appear to be a nonlinear 0--1 optimization problem because relevance is decayed by latency and combined with retrieval cost. In the base formulation, however, each candidate's utility is fixed before selection, reducing the problem to a linear objective with structured constraints. As a result, it is solvable exactly in polynomial time; NP-hardness appears only under additional constraints such as a hard egress budget.

\subsection{The base problem is polynomial}

For each candidate, define its \emph{net utility}
\begin{equation}
u_i \;=\; \mathrm{rel}_i\, e^{-\gamma_q\,\mathrm{lat}_i(s)} \;-\; \beta\,\mathrm{cost}_i .
\label{eq:netutil}
\end{equation}
For a given query, all terms in $u_i$ are known before selection. Thus,
Eq.~\eqref{eq:objective} becomes linear in 
$x$:
$\max_x \sum_i u_i x_i$. The exponential term only affects the fixed coefficient
$u_i$; it does not make the optimization nonlinear.

The regional caps in Eq.~\eqref{eq:regional} define a partition matroid because each document
belongs to exactly one region. The global top-$k$ constraint in Eq.~\eqref{eq:cardinality} is a
truncation of that matroid. Therefore, a linear objective can be optimized exactly by greedy
selection.

\begin{quote}
\textbf{Proposition 1.} \emph{Consider candidates with $u_i>0$ and sort them in decreasing order
of $u_i$. Scan them in that order and select a candidate whenever doing so respects both its
regional cap $C_{r(i)}$ and the global limit $k$. Stop once $k$ documents are selected or no
positive-utility candidate remains. This algorithm is optimal and runs in $O(N\log N)$ time.}
\end{quote}

The runtime is dominated by sorting; the subsequent scan is linear. Because
Eq.~\eqref{eq:cardinality} is an inequality, an optimal solution never includes a candidate with
$u_i<0$. GRaSP may therefore select fewer than $k$ documents, or none, when latency and cost make
all remaining candidates unattractive.

The base GRaSP problem is therefore not NP-hard. It admits a simple exact algorithm suitable for
the retrieval path, which we validate against an integer-program oracle in
Section~\ref{sec:eval}.

\subsection{Where hardness actually enters}

Hardness appears when the model includes an additive resource constraint, such as a hard egress
budget
\begin{equation}
\sum_{i=1}^N x_i\,\mathrm{cost}_i \;\le\; B .
\label{eq:budget}
\end{equation}
This constraint couples candidate choices through their costs. Maximizing
$\sum_i u_i x_i$ under Eq.~\eqref{eq:budget} contains 0--1 knapsack as a special case and is
therefore NP-hard.

More complex compliance rules that introduce dependencies across regions or documents can likewise
break the matroid structure. Approximation or Lagrangian methods are therefore needed only for
these harder variants, not for the base GRaSP formulation. In our evaluation, an exact
integer-program solver serves as an offline oracle for measuring their optimality gap.

\section{Algorithms}
\label{sec:algorithms}

For the base GRaSP problem, Proposition~1 gives an exact $O(N\log N)$ solution: sort candidates by
net utility and greedily select them while respecting the global and regional caps. We use this
\emph{cap-aware greedy} as the default policy. We consider two extensions: a Lagrangian variant
for harder budgeted or compliance-constrained formulations, and a simpler latency-decay function
for high-throughput serving.

\subsection{Lagrangian-greedy for constrained variants}

When additional constraints break the matroid structure, we use regional multipliers
$\lambda_r \ge 0$ and rank candidates by
\begin{equation}
\label{eq:lag-score}
\Delta_i \;=\; \mathrm{rel}_i\,e^{-\gamma_q\,\mathrm{lat}_i(s)} \;-\; \beta\,\mathrm{cost}_i
\;-\; \lambda_{r(i)}.
\end{equation}

Candidates are ranked by $\Delta_i$ and selected under the global budget $k$. The multipliers are
then updated using a subgradient step: regions that exceed their desired capacity receive a larger
penalty, while penalties are reduced for under-used regions.

For the base problem, this machinery is unnecessary because the direct greedy algorithm is already
optimal. We use the Lagrangian form only for extensions with hard budgets or more complex
compliance constraints, where it provides a lightweight and parallelizable approximation.

\subsection{Rational decay for fast scoring}

For very high-throughput settings, we approximate the exponential latency decay with
\begin{equation}
\label{eq:pade}
e^{-\gamma_q\,\mathrm{lat}_i(s)} \;\approx\; \frac{1}{1+\gamma_q\,\mathrm{lat}_i(s)},
\end{equation}
which gives the score
\begin{equation}
\label{eq:fastscore}
\tilde{\Delta}_i \;=\; \frac{\mathrm{rel}_i}{1+\gamma_q\,\mathrm{lat}_i(s)}
\;-\; \beta\,\mathrm{cost}_i .
\end{equation}

We then apply the same cap-aware selection procedure using $\tilde{\Delta}_i$. The rational form is
cheaper to evaluate and slightly under-penalizes large latencies compared with the exponential
decay. As a result, it tends to preserve slightly more recall at the cost of a somewhat heavier
latency tail.

\paragraph{Complexity}
Both variants require $O(N\log N)$ time per query when implemented with sorting, or $O(N)$ with
selection-based implementations. For candidate pools of a few thousand documents, this overhead
is small relative to retrieval and LLM inference, as shown in Section~\ref{sec:overhead}.

\section{Evaluation}
\label{sec:eval}

Our evaluation answers four questions. (Q1) Does latency-aware selection reduce tail latency and cost, and what retrieval-quality trade-off does it introduce? (Q2) How does this trade-off depend on data placement? (Q3) Can replication recover quality lost to latency awareness? (Q4) Is GRaSP close to optimal while remaining lightweight and stable enough for online serving?

\subsection{Experimental methodology and assumptions}
\label{sec:methodology}

We first describe the testbed and modeling choices used throughout the evaluation. Several choices---including relevance normalization, pricing, and placement---are not fixed by the GRaSP formulation itself, so we test their sensitivity separately.

\textbf{Testbed:}
We use a five-node CloudLab~\cite{CloudLab} cluster (Utah \texttt{m510}: 8-core Xeon-D, 64\,GB RAM, NVMe, 10\,GbE). One node acts as the query coordinator (the source region $s$), while four nodes host document payloads in separate logical regions. We emulate wide-area latency using Linux \texttt{tc netem}, with round-trip times of $\{1, 90, 210, 115 \}$ ms for the local, \texttt{eu-central}, \texttt{ap-southeast}, and \texttt{sa-east} regions, respectively. These values are consistent with representative public AWS inter-region latency measurements~\cite{rttsource}. We use persistent connections and pin static ARP entries so that connection setup does not perturb the emulated latency matrix.

\textbf{Corpus and retriever:}
Our main corpus is BEIR Natural Questions (NQ)~\cite{kwiatkowski2019natural,thakur2021beir}, containing 2{,}681{,}468 passages and 3{,}452 test queries with relevance judgments. To test generalization across domains, we additionally evaluate SciFact, NFCorpus, and ArguAna~\cite{thakur2021beir}, covering scientific claim verification, biomedical retrieval, and argument retrieval.

Passages are embedded with a sentence-transformer model~\cite{reimers2019sentence} and indexed using exact inner-product FAISS~\cite{johnson2019billion}. Each query produces a top-200 candidate pool, or the full corpus when the dataset is smaller. GRaSP operates only on this supplied candidate pool: it does not change the retriever, embedding model, or index. Stronger upstream retrievers may therefore shift absolute recall, but the selection mechanism itself is retriever-agnostic.

\textbf{End-task generation:}
Retrieval metrics do not fully capture answer quality, so we also evaluate an end-to-end QA task on NQ. We condition a generator~\cite{chung2024scaling} on the context selected by each policy and report exact match (EM) and token-level F1 against short gold answers. Gold answers are obtained by joining NQ-Open~\cite{kwiatkowski2019natural} with the BEIR NQ queries by question text (Section~\ref{sec:endtask}).

\textbf{Relevance normalization:}
GRaSP assumes $\mathrm{rel}_i\in[0,1]$ but does not prescribe how raw retriever scores should be mapped into this range. We use per-query min--max normalization by default. Since the latency decay multiplies $\mathrm{rel}_i$, the score distribution affects the numerical meaning of $\gamma_q$. We therefore repeat the main experiment with softmax normalization in Section~\ref{sec:robustness}.

\textbf{Cost model:}
We compute per-document retrieval cost as passage size multiplied by a representative regional egress price
(\{0.01, 0.02, 0.09, 0.06\} USD/GB), consistent with published AWS data-transfer rates~\cite{egress}. Costs are normalized within each candidate pool so that $\beta$ operates on a scale comparable to latency-adjusted relevance. Batching discounts and tiered pricing are outside the scope of the model.

In the default price map, distant regions also tend to be more expensive, so latency and cost are correlated. To isolate the effect of $\beta$ from $\gamma_q$, we also evaluate a decorrelated map in which a relatively near region is the most expensive (Section~\ref{sec:beta}).

\textbf{Placement:}
We study two placement regimes. Under \emph{uniform placement}, documents are hashed across the four regions. Relevant candidates are therefore spread across regions, making nearby alternatives relatively common.

Under \emph{topical placement}, document embeddings are clustered into four groups using $k$-means and each cluster is mapped to a region. This approximates deployments in which related content is co-located because of organizational ownership, jurisdiction, or data-residency rules. The average concentration of a query's top-5 candidates in its dominant region is $0.89$ under topical placement, compared with approximately $0.40$ under uniform placement. Topical placement therefore creates the setting in which the most relevant content is often concentrated in one, potentially remote, region.

\textbf{Policies:}
We compare \emph{Relevance-only}, the standard top-$k$ policy; \emph{Nearest-$k$}, a latency-only baseline; \emph{GRaSP-greedy}, the direct cap-aware greedy algorithm from Proposition~1; \emph{GRaSP-rational}, which uses Eq.~\eqref{eq:fastscore}; and \emph{OPT}, an exact integer-program solution of Eq.~\eqref{eq:objective} used offline as an optimality oracle. Unless stated otherwise, $k=5$.

Nearest-$k$ is intentionally simple, so we additionally evaluate \emph{Rel@LatencyCap}. Given a threshold $L$, it removes candidates with latency above $L$ and selects the highest-relevance feasible documents among the rest. Sweeping $L$ over the regional latencies gives this baseline its own recall--latency frontier and tests whether GRaSP provides more than a hard latency filter.

\textbf{Metrics:}
Offline, we report optimality gap relative to OPT, expected worst-case fetch latency, egress cost, recall@5, nDCG@5, and the fraction of queries producing an empty selection. We define the mean per-query maximum fetch latency as
\begin{equation}
\label{eq:lmax}
\bar{L}_{\max} \;=\; \frac{1}{|Q|}\sum_{q\in Q}\ \max_{d\in S_q}\
\mathrm{lat}\!\big(d,s_q\big),
\end{equation}
with $\max_{d\in\emptyset}(\cdot)=0$ for an empty selection.

Because selected passages are fetched in parallel over persistent connections, retrieval completes when the slowest selected passage arrives. $\bar L_{\max}$ therefore provides a natural offline approximation of fetch latency, although it does not include protocol, scheduling, or response-processing overhead.

For online measurements, we use an open-loop asynchronous load generator with persistent per-region connections and report end-to-end retrieval latency percentiles together with per-query selection overhead.

\textbf{Threats to validity:}
Our experiments use a single CloudLab cluster with an emulated WAN rather than a live multi-region deployment. The RTT matrix is a point estimate and does not capture burstiness, routing changes, or diurnal variation. Pricing and relevance normalization are also modeling choices rather than properties of GRaSP itself. We address these limitations by testing alternative normalization and cost maps, evaluating both uniform and topical placement, reporting three-seed variation, and repeating the main comparison across four datasets. Our conclusions focus on relative trade-offs and placement dependence rather than absolute latency for a specific cloud deployment. A live multi-region evaluation remains future work.

\subsection{Q1/Q2: Latency, cost, quality, and placement}
\label{sec:frontier}

We first study how $\gamma_q$ changes GRaSP's behavior under topical placement. Table~\ref{tab:sweep} shows representative points from the sweep. As $\gamma_q$ increases, remote documents receive a larger penalty, shifting the selected context toward nearby regions. Latency and egress cost therefore fall, while recall gradually decreases.

At $\gamma_q=0.003$, GRaSP-greedy reduces expected worst-case fetch latency from $122.1$ to $97.4$\,ms while retaining recall $0.506$ versus $0.534$ for relevance-only. At $\gamma_q=0.010$, latency falls to $69.5$\,ms with recall $0.419$. At $\gamma_q=0.030$, GRaSP approaches the latency-only extreme, reaching $47.2$\,ms with recall $0.331$. Thus, $\gamma_q$ provides a continuous operating knob between relevance and locality.

GRaSP-greedy matches OPT throughout the sweep, confirming Proposition~1 empirically. GRaSP-rational also remains close to optimal at low and moderate $\gamma_q$, but diverges at the largest setting. This matches the approximation's expected behavior: rational decay penalizes very large latencies less aggressively than the exponential, preserving somewhat more recall but producing a heavier latency tail.

\begin{table}[t]
\centering
\caption{Topical placement, $\beta{=}0$, $k{=}5$. Optimality gap vs.\ OPT, expected worst-case
fetch latency (ms), egress cost ($\mu$USD per query), recall@5, and nDCG@5.
Values are means over 3{,}452 queries. (Representative rows; full sweep in the artifact.)}
\label{tab:sweep}
\small
\begin{tabular}{llccccc}
\toprule
$\gamma_q$ & Policy & Gap & Lat.\ (ms) & Cost & Recall & nDCG \\
\midrule
0.003 & Rel-only     & 0.92 & 122.1 & 0.102 & 0.534 & 0.400 \\
0.003 & Nearest-$k$  & 0.76 &  42.3 & 0.044 & 0.292 & 0.229 \\
0.003 & GRaSP-greedy & 1.00 &  97.4 & 0.084 & 0.506 & 0.384 \\
0.003 & GRaSP-rat.\  & 1.00 & 102.3 & 0.088 & 0.516 & 0.391 \\
\midrule
0.010 & Rel-only     & 0.85 & 122.1 & 0.102 & 0.534 & 0.400 \\
0.010 & Nearest-$k$  & 0.87 &  42.3 & 0.044 & 0.292 & 0.229 \\
0.010 & GRaSP-greedy & 1.00 &  69.5 & 0.065 & 0.419 & 0.323 \\
0.010 & GRaSP-rat.\  & 0.98 &  83.3 & 0.076 & 0.466 & 0.357 \\
\midrule
0.030 & Rel-only     & 0.59 & 122.1 & 0.102 & 0.534 & 0.400 \\
0.030 & Nearest-$k$  & 0.97 &  42.3 & 0.044 & 0.292 & 0.229 \\
0.030 & GRaSP-greedy & 1.00 &  47.2 & 0.049 & 0.331 & 0.258 \\
0.030 & GRaSP-rat.\  & 0.91 &  68.5 & 0.065 & 0.413 & 0.320 \\
\bottomrule
\end{tabular}
\end{table}

Absolute egress costs are small per query, but their relative reduction matters at scale. At $\gamma_q=0.030$, for example, GRaSP reduces cost from $0.102$ to $0.049\,\mu$USD/query, roughly a $2\times$ reduction. At billions of RAG queries per day, differences of this size can accumulate to hundreds or thousands of dollars in daily transfer cost.

\begin{figure*}[t]
\centering
\includegraphics[width=0.85\linewidth]{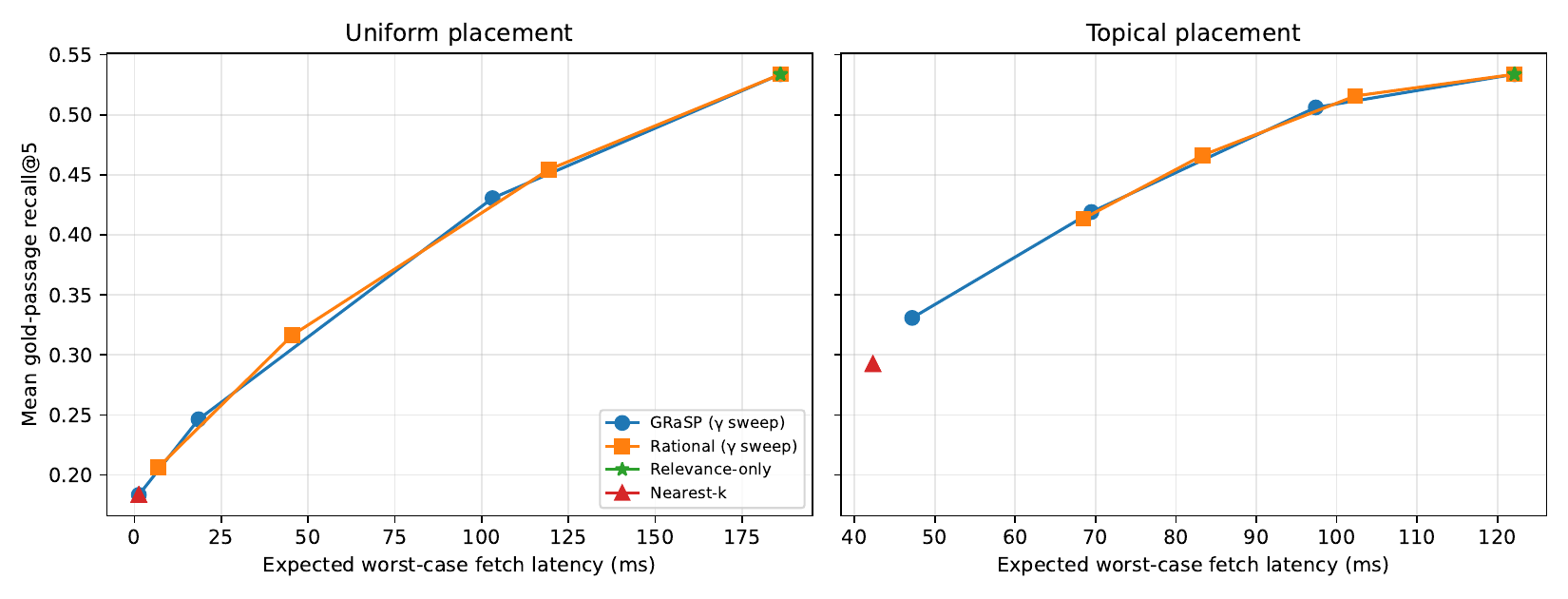}
\caption{Recall@5 versus expected worst-case fetch latency as $\gamma_q$ varies
($\beta{=}0$). Left: uniform placement---GRaSP reduces latency at negligible recall cost. Right:
topical placement---GRaSP traces a Pareto frontier between retrieval quality and latency.}
\label{fig:frontier}
\end{figure*}

Figure~\ref{fig:frontier} contrasts the two placement regimes and captures the paper's central result. Under \emph{uniform placement}, the $\gamma_q$ sweep remains close to the top-left of the recall--latency plane: latency falls substantially before recall begins to decline. Because relevant documents are distributed across regions, GRaSP can often replace a remote passage with a similarly relevant nearby one.

Under \emph{topical placement}, the frontier bends outward. A query's most relevant candidates tend to reside together, so moving toward nearby regions increasingly requires accepting less relevant passages. GRaSP still exposes a smooth recall--latency trade-off, but the gain is no longer free.

The lesson is that latency-aware selection can exploit locality only when good nearby alternatives exist. If the relevant content is concentrated in a remote region, no choice of $\gamma_q$ can remove the remote fetch without sacrificing quality. This placement dependence is therefore a property of the available candidate set, not a limitation of the optimization algorithm.

\textbf{A stronger latency-aware baseline:}
Nearest-$k$ represents the latency-only extreme and is intentionally weak. A more realistic practitioner might impose a latency threshold and then maximize relevance within it. Rel@LatencyCap models this approach.

We sweep $L\in\{1,90,115,210\}$\,ms under NQ topical placement. At $L=115$\,ms, the baseline achieves recall $0.432$ at $79.8$\,ms. GRaSP provides nearby operating points with recall $0.419$ at $69.5$\,ms and $0.506$ at $97.4$\,ms. More generally, its $\gamma_q$ sweep produces a better interior trade-off because latency is penalized continuously rather than treated as an all-or-nothing cutoff.

\begin{table}[t]
\centering
\caption{Rel@LatencyCap baseline on NQ topical placement: discard candidates with
$\mathrm{rtt}>L$, then select the highest-relevance feasible set.}
\label{tab:latcap}
\small
\begin{tabular}{cccc}
\toprule
$L$ (ms) & Recall & nDCG & Mean max-RTT (ms) \\
\midrule
1   & 0.000 & 0.000 &   0.0 \\
90  & 0.260 & 0.198 &  52.4 \\
115 & 0.432 & 0.324 &  79.8 \\
210 & 0.534 & 0.400 & 122.1 \\
\bottomrule
\end{tabular}
\end{table}

The advantage appears in the operating range where tuning is most useful: a hard threshold cannot distinguish a relevant candidate just beyond $L$ from a weak candidate just below it, whereas GRaSP trades relevance and latency jointly for each document.

\subsection{Online tail latency}
\label{sec:online}

\begin{figure}[t]
\centering
\includegraphics[width=\linewidth]{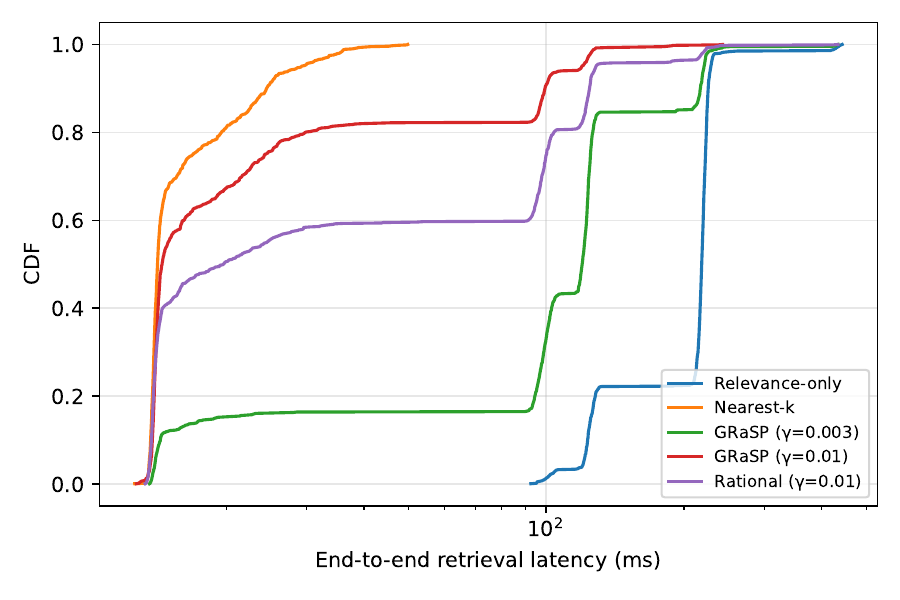}
\caption{Measured end-to-end retrieval latency (CDF, log-$x$) under topical placement at
20~queries/s. Increasing $\gamma_q$ shifts GRaSP's latency distribution left of relevance-only.}
\label{fig:cdf}
\end{figure}

We next measure actual end-to-end retrieval latency under topical placement at 20~queries/s. Figure~\ref{fig:cdf} shows that online measurements preserve the offline ordering: relevance-only is slowest, nearest-$k$ is fastest, and GRaSP lies between them, with $\gamma_q$ controlling the operating point.

Absolute online latency is higher than the offline $\bar L_{\max}$ estimate, and the difference is policy-dependent. The offline metric models only the slowest selected document's network latency, whereas the online path also includes HTTP request/response handling and coordination across parallel fetches. GRaSP often selects a mixture of near and far documents, making this overhead more visible than for policies whose requests are concentrated in one latency class.

Thus, $\bar L_{\max}$ should be treated as a surrogate for comparing policies rather than an exact prediction of serving latency. Importantly, it predicts both the relative ordering and the direction of GRaSP's improvement. Across three seeds, median latency varies by only $\pm0.1$--$0.3$\,ms, while P99 varies by approximately $\pm8$--$10$\,ms. At $\gamma_q=0.01$, GRaSP-greedy reduces median latency by about $20\%$ and P90 by about $41\%$ relative to relevance-only while retaining approximately $78\%$ of its recall. GRaSP-rational behaves similarly near the median but exhibits a heavier P90/P99 tail, consistent with its weaker penalty on distant candidates.

\subsection{Q3: Replication as a design lever}
\label{sec:replication}

The topical-placement result exposes a limitation of selection alone: if all highly relevant content is remote, GRaSP must either fetch it or sacrifice relevance. Placement, however, need not remain fixed.

\begin{figure}[t]
\centering
\includegraphics[width=\linewidth]{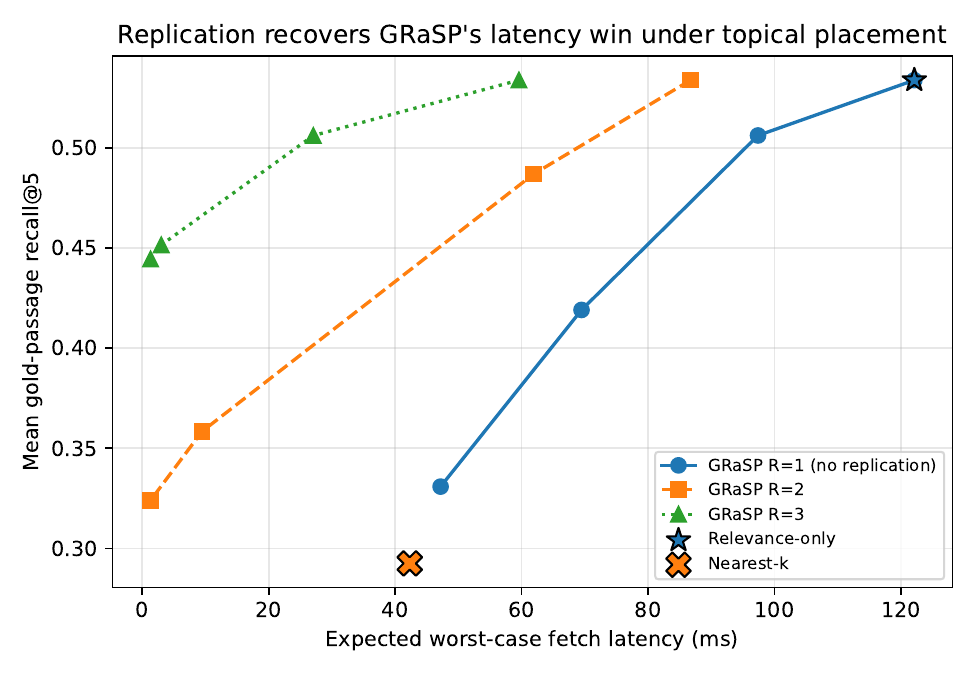}
\caption{Replication improves GRaSP's recall--latency frontier under topical placement.
Higher replication factors move the frontier toward the uniform-placement regime.}
\label{fig:replication}
\end{figure}

We therefore replicate each document to its topical home plus $R-1$ additional regions and fetch from the lowest-latency replica. Figure~\ref{fig:replication} shows the resulting frontiers for $R\in\{1,2,3\}$.

As $R$ increases, the frontier moves toward the uniform-placement regime. At comparable retrieval quality, replicating content to two or three regions roughly \emph{halves} expected fetch latency and egress cost because a relevant document is more likely to have a nearby, cheaper copy.

This suggests a natural co-design rule: replicate frequently accessed content, then let GRaSP exploit whichever copy is most attractive at query time. GRaSP therefore complements caching and placement mechanisms rather than replacing them.

Our replication model is intentionally optimistic: it assumes the lowest-latency replica is always reachable and does not model replica consistency, update traffic, or storage overhead. The resulting curves should therefore be viewed as an upper bound on the achievable latency reduction.

\subsection{Residency caps}
\label{sec:caps}

We next activate the per-region constraints by setting $C_r=2$ under topical placement. Since relevant content is often concentrated in one region, the cap forces every policy to obtain part of its context from other regions.

Table~\ref{tab:caps} shows the effect at $\gamma_q=0.01$. For relevance-only, the cap raises expected latency from $122.1$ to $175.3$\,ms and reduces recall from $0.534$ to $0.392$. After exhausting the dominant region, relevance-only simply selects the highest-ranked feasible documents without considering their location.

GRaSP also loses recall and gains latency because the feasible set itself has become worse, but it continues to match OPT. The cap changes what is achievable; GRaSP still finds the best feasible selection under that constraint.

\begin{table}[t]
\centering
\caption{Effect of a residency cap ($C_r{=}2$) under topical placement, $\gamma_q{=}0.01$,
$\beta{=}0$. GRaSP tracks the exact optimum.}
\label{tab:caps}
\small
\begin{tabular}{llccc}
\toprule
Setting & Policy & Gap & Lat.\ (ms) & Recall \\
\midrule
Uncapped & Rel-only     & 0.85 & 122.1 & 0.534 \\
Uncapped & GRaSP-greedy & 1.00 &  69.5 & 0.419 \\
\midrule
$C_r{=}2$ & Rel-only     & 0.97 & 175.3 & 0.392 \\
$C_r{=}2$ & GRaSP-greedy & 1.00 & 163.3 & 0.374 \\
\bottomrule
\end{tabular}
\end{table}

A related behavior appears under strong joint latency and cost pressure. Some candidates can then have negative net utility. Relevance-only and nearest-$k$ may still select them because their policies consider only relevance or locality. GRaSP instead selects fewer than $k$ documents, or none, when another retrieval would lower the objective. This is the abstention behavior described in Section~\ref{sec:complexity}.

\subsection{Q4: Optimality, overhead, and load stability}
\label{sec:overhead}

For GRaSP to be useful on the serving path, the selection benefit must not be offset by optimization overhead. We therefore examine both optimality and performance as query load increases.

GRaSP-greedy achieves an optimality gap of $1.00$ across the evaluated base-problem settings, empirically confirming Proposition~1. GRaSP-rational remains above $0.98$ over the low-to-moderate $\gamma_q$ range and diverges only near the aggressive local-only regime. In contrast, solving OPT with a general integer-program solver adds tens of milliseconds per query and is unsuitable for online serving.

Table~\ref{tab:qps} shows behavior from 20 to 200 queries/s. Median latency rises only modestly with load, and GRaSP remains faster than relevance-only at every tested rate. Selection itself costs only hundreds of microseconds per query, far below the wide-area fetch latency.

\begin{table}[t]
\centering
\caption{Load stability under topical placement ($\gamma_q{=}0.01$, $\beta{=}0$). End-to-end
latency percentiles (ms) and selection overhead ($\mu$s), 1{,}000 queries per point.}
\label{tab:qps}
\small
\begin{tabular}{llcccc}
\toprule
QPS & Policy & P50 & P90 & P99 & Sel.\ ($\mu$s) \\
\midrule
20  & Rel-only     & 130.5 & 228.3 & 247.0 & 125 \\
20  & GRaSP-greedy & 105.1 & 135.4 & 248.4 & 425 \\
50  & Rel-only     & 131.3 & 228.4 & 426.9 & 125 \\
50  & GRaSP-greedy & 105.8 & 139.7 & 290.9 & 426 \\
100 & Rel-only     & 132.9 & 229.7 & 433.4 & 118 \\
100 & GRaSP-greedy & 106.6 & 146.7 & 425.8 & 419 \\
200 & Rel-only     & 138.8 & 232.6 & 444.0 &  91 \\
200 & GRaSP-greedy & 111.3 & 173.7 & 294.1 & 271 \\
\bottomrule
\end{tabular}
\end{table}

At 200 queries/s, GRaSP's P50 remains $111.3$\,ms compared with $138.8$\,ms for relevance-only, while selection overhead is only $271\,\mu$s. Even at the highest tested load, serving latency is dominated by the wide-area content fetch rather than the GRaSP layer.

\subsection{Robustness to relevance normalization}
\label{sec:robustness}

Because $\gamma_q$ acts on normalized relevance, its numerical value depends on the score scale. To test whether our conclusions are an artifact of min--max normalization, we repeat the topical-placement sweep using softmax-normalized scores.

\begin{figure}[t]
\centering
\includegraphics[width=\linewidth]{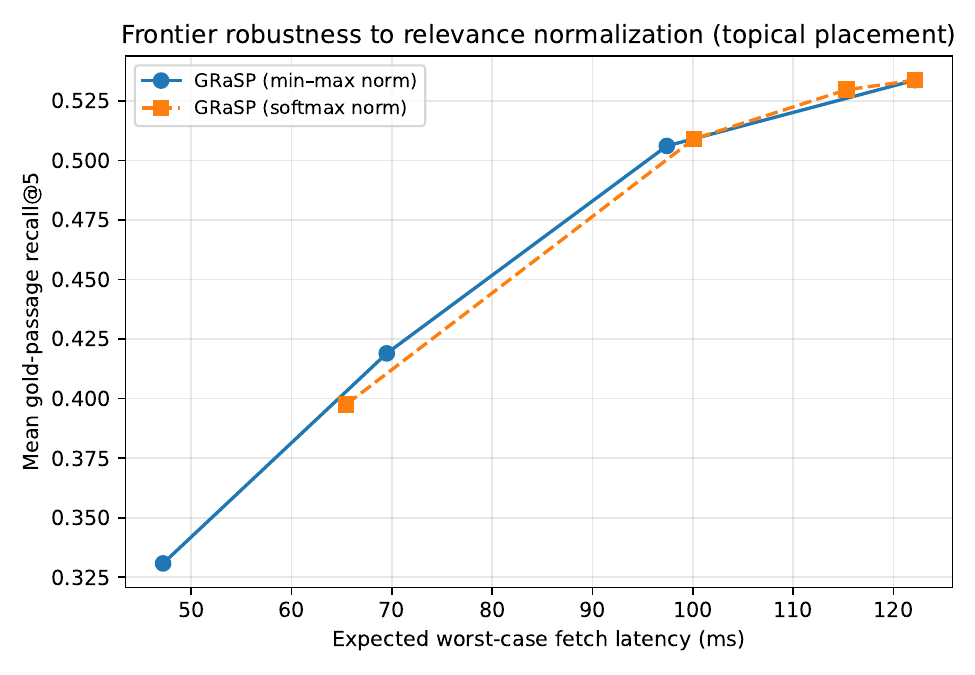}
\caption{Frontier robustness to relevance normalization under topical placement. Min--max and
softmax normalization shift individual operating points but preserve the overall recall--latency
trade-off.}
\label{fig:robustness}
\end{figure}

Figure~\ref{fig:robustness} shows that normalization shifts the $\gamma_q$ value corresponding to a particular operating point, but does not change the qualitative shape of the recall--latency frontier. GRaSP-greedy continues to match OPT, and the same placement-dependent trade-off remains.

\begin{table}[b]
\centering
\caption{Decorrelated pricing (near region most expensive), topical placement,
$\gamma_q{=}0.01$. Increasing $\beta$ trades latency and relevance for lower egress cost.}
\label{tab:beta}
\small
\begin{tabular}{cccc}
\toprule
$\beta$ & Latency (ms) & Cost ($\mu$USD) & Recall \\
\midrule
0.0 & 69.5 & 0.095 & 0.419 \\
0.5 & 70.2 & 0.051 & 0.339 \\
1.0 & 70.8 & 0.036 & 0.279 \\
2.0 & 74.3 & 0.024 & 0.210 \\
\bottomrule
\end{tabular}
\end{table}

We therefore use min--max normalization as the default, while treating $\gamma_q$ as a parameter that should be calibrated to the relevance-score distribution of the deployed retriever.

\subsection{The cost knob $\beta$ is independent of the latency knob $\gamma_q$}
\label{sec:beta}

Under the default price map, distant regions are also relatively expensive. Increasing $\beta$ therefore tends to have the same directional effect as increasing $\gamma_q$, making the two controls difficult to separate.

To isolate the cost effect, we construct a decorrelated price map in which the 90\,ms region is the most expensive. We then hold $\gamma_q=0.01$ fixed and sweep $\beta$.

\begin{figure}[t]
\centering
\includegraphics[width=\linewidth]{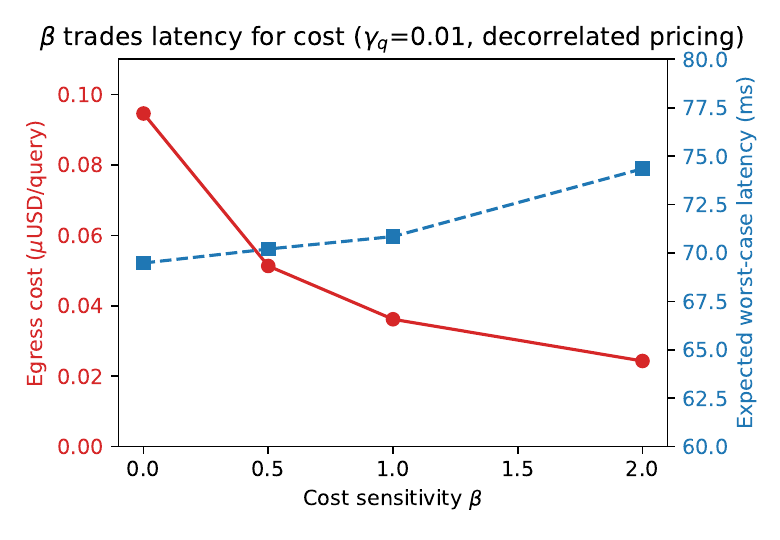}
\caption{Under decorrelated pricing ($\gamma_q{=}0.01$), increasing $\beta$ reduces egress cost
while latency rises. The latency and cost knobs therefore control distinct trade-offs.}
\label{fig:beta}
\end{figure}

As Table~\ref{tab:beta} and Figure~\ref{fig:beta} show, increasing $\beta$ from $0$ to $2$ reduces egress cost from $0.095$ to $0.024\,\mu$USD/query, a $74\%$ reduction. Latency simultaneously rises from $69.5$ to $74.3$\,ms because GRaSP moves away from the nearby but expensive region toward cheaper, more distant documents.

The opposite movement of cost and latency confirms that $\beta$ and $\gamma_q$ are genuinely independent controls. Applications can therefore prioritize interactive latency and monetary cost separately.

\subsection{Generalization across datasets}
\label{sec:generalization}

We repeat the core experiment on SciFact, NFCorpus, and ArguAna. Their average top-5 candidate concentrations under topical placement are $0.83$, $0.72$, and $0.92$, respectively, compared with $0.89$ for NQ. All four datasets therefore exhibit geographic concentration, although its strength varies.

Table~\ref{tab:general} reports the main operating point at $\gamma_q=0.01$ and $\beta=0$. Two patterns hold across all datasets. First, GRaSP-greedy achieves an optimality gap of $1.000$, confirming the exactness of Proposition~1 across domains. Second, recall follows the expected ordering
relevance-only $\ge$ GRaSP $\ge$ nearest-$k$, while GRaSP substantially reduces expected fetch latency.

\begin{table}[t]
\centering
\caption{Generalization across datasets at $\gamma_q{=}0.01$, $\beta{=}0$, topical placement.
GRaSP-greedy matches the exact optimum on every dataset.}
\label{tab:general}
\small
\begin{tabular}{llcccc}
\toprule
Dataset & Policy & Gap & Lat.\ (ms) & Recall & nDCG \\
\midrule
NQ       & Rel-only     & 0.85 & 122.1 & 0.534 & 0.400 \\
& GRaSP-greedy & 1.00 &  69.5 & 0.419 & 0.323 \\
& Nearest-$k$  & 0.87 &  42.3 & 0.292 & 0.229 \\
\midrule
SciFact  & Rel-only     & 0.82 & 119.6 & 0.738 & 0.629 \\
& GRaSP-greedy & 1.00 &  61.6 & 0.560 & 0.489 \\
& Nearest-$k$  & 0.85 &  13.3 & 0.276 & 0.240 \\
\midrule
NFCorpus & Rel-only     & 0.77 & 141.4 & 0.120 & 0.350 \\
& GRaSP-greedy & 1.00 &  45.4 & 0.082 & 0.287 \\
& Nearest-$k$  & 0.92 &  10.1 & 0.051 & 0.210 \\
\midrule
ArguAna  & Rel-only     & 0.87 & 109.2 & 0.575 & 0.308 \\
& GRaSP-greedy & 1.00 &  73.9 & 0.459 & 0.255 \\
& Nearest-$k$  & 0.80 &  30.7 & 0.262 & 0.150 \\
\bottomrule
\end{tabular}
\end{table}

The exact trade-off varies by corpus because different datasets provide different numbers and qualities of nearby substitutes for remote relevant passages. This variation reinforces the placement result: GRaSP's benefit depends on the local availability of useful alternatives, not on a particular dataset.

\subsection{End-task answer quality}
\label{sec:endtask}

Finally, we examine whether the retrieval-quality loss translates into a comparable loss in generated answer quality. We evaluate 500 NQ queries and condition the generator on the passages selected by each policy.

Table~\ref{tab:endtask} shows that GRaSP retains approximately $81\%$ of relevance-only's EM and $87\%$ of its F1 while reducing context-fetch latency by approximately $43\%$. It also substantially outperforms nearest-$k$ on both answer-quality metrics.

\begin{table}[t]
\centering
\caption{End-task NQ answer quality (500-query subset, $\gamma_q{=}0.01$, $\beta{=}0$, topical
placement). GRaSP preserves most answer quality while reducing context-fetch latency.}
\label{tab:endtask}
\small
\begin{tabular}{lccc}
\toprule
Policy & EM & F1 & Context lat.\ (ms) \\
\midrule
Relevance-only & 0.320 & 0.415 & 122.1 \\
GRaSP-greedy   & 0.258 & 0.362 &  69.5 \\
Nearest-$k$    & 0.180 & 0.257 &  42.3 \\
\bottomrule
\end{tabular}
\end{table}

Interestingly, answer quality falls less than retrieval recall. At this operating point, GRaSP retains about $78\%$ of relevance-only's recall but $81\%$ of its EM and $87\%$ of its F1. Missing a gold passage therefore does not always prevent the generator from answering correctly; other selected passages may contain sufficient or overlapping evidence. Recall@5 is, if anything, pessimistic about the downstream effect of GRaSP's retrieval trade-off. Paired bootstrap tests with 10{,}000 resamples show the differences are statistically significant. Relative to relevance-only, GRaSP's EM decreases by $0.062$ (95\% CI $[0.032,0.092]$), while context-fetch latency falls by $53$\,ms (CI $[47,60]$). Compared with nearest-$k$, GRaSP improves EM by $0.078$ (CI $[0.052,0.106]$) and F1 by $0.104$ (CI $[0.076,0.133]$).

Together, these results show that GRaSP's latency savings persist in the online retrieval path and end-to-end QA setting, while the loss in answer quality is smaller than retrieval recall alone suggests.

\section{Discussion and Limitations}
The optimization itself is deliberately simple; our contribution is to identify and empirically characterize the conditions under which location-aware context selection helps rather than designing a sophisticated solver.
  
\textbf{Drop-in adoption:}
GRaSP is designed to fit into an existing RAG pipeline with minimal changes. It requires only a latency estimate for each candidate's hosting region---from a static RTT table, recent measurements, or a learned predictor---and, optionally, a regional cost map. After an upstream retriever or re-ranker produces a candidate pool, GRaSP re-scores them and enforces the configured global and regional limits before payloads are fetched. It therefore requires no retraining, embedding changes, or re-indexing.

\textbf{When GRaSP helps:}
Our results show that the value of geo-aware selection depends primarily on where useful alternatives are located. When relevant content is uniformly distributed or replicated across regions, GRaSP can reduce latency and egress cost with little loss in retrieval quality. When relevant content is concentrated in a remote region, as may happen under residency or ownership constraints, locality and relevance are in genuine tension. In this regime, GRaSP does not eliminate the trade-off; instead, it exposes a controllable Pareto frontier through $\gamma_q$ and $\beta$.

This also suggests that placement and query-time selection should be considered together. Replicating frequently accessed content into even a small number of additional regions increases the chance that GRaSP can choose a nearby high-quality passage. Our replication experiments show that this can move the system back toward the regime in which latency and cost reductions are obtained with little quality loss.

\textbf{Limitations:}
First, the current formulation assumes a point estimate of network latency. Real wide-area networks exhibit burstiness, asymmetry, congestion, and time-dependent behavior. These effects could be incorporated using percentile latency estimates, moving averages, or time-of-day profiles, but we do not evaluate those variants here.

Second, our cost model is additive at the document level. Real cloud pricing may include batching effects, volume discounts, minimum charges, or tiered pricing. Some of these can be incorporated into the objective or added as linear constraints; a tight aggregate budget leads to the budgeted variant discussed in Section~\ref{sec:complexity}, where exact greedy selection no longer applies.

Third, our systems evaluation uses an emulated WAN on a single CloudLab cluster rather than a live multi-region deployment. We also use one retrieval setup and one primary large-scale corpus, supplemented by three smaller BEIR datasets. These experiments are sufficient to expose the placement-dependent behavior we study, but they do not capture all production effects, such as routing variation, regional failures, provider-specific networking, or interactions with different retrievers and corpora.

Fourth, the replication experiments model an optimistic setting in which the lowest-latency replica is available and can be selected without additional coordination. We do not account for replica consistency, update propagation, storage overhead, or the cost of maintaining those copies. The measured replication gains should therefore be interpreted as an upper bound on what a production system could achieve.

\textbf{Outlook:}
Several extensions follow naturally from this work. The parameters $\gamma_q$, $\beta$, and even $k$ could be chosen dynamically from workload characteristics or learned from user feedback. Placement and replication could also be optimized jointly with query-time selection, allowing the system to decide both where content should live and which copy should serve each request. Finally, the same formulation could be extended beyond independent passage retrieval to account for multi-hop or graph-based reasoning, where the location and cost of one retrieval may influence which document should be fetched next.

\section{Conclusion}

We presented GRaSP, a latency- and cost-aware re-scoring layer for geo-distributed RAG. GRaSP adds location awareness to retrieval without retraining or re-indexing, and its base formulation remains simple: the selection problem is solvable exactly with an $O(N\log N)$ greedy algorithm, while NP-hardness arises only in extensions with hard additive budgets.

Our evaluation on a five-node CloudLab testbed and 2.68M passages shows that the benefit of geo-aware retrieval depends strongly on data placement. When relevant content is uniformly distributed or replicated, GRaSP reduces latency and cost with little loss in retrieval quality. When relevant content is concentrated in remote regions, latency and relevance are in genuine tension, and GRaSP exposes a tunable trade-off between them. Replication provides a practical way to reduce this tension by bringing useful alternatives closer to the serving region.

GRaSP adds sub-millisecond selection overhead, matches the exact solution for the base problem, and remains stable up to 200~queries/s. More broadly, our results show that retrieval quality alone is not sufficient for geo-distributed RAG: where relevant content resides can be just as important as how relevant it is.

\bibliographystyle{ACM-Reference-Format}
\balance
\bibliography{Bib}

\end{document}